\documentclass[onecolumn,authoryear]{aastex63}
\usepackage{graphicx}	
\usepackage{amsmath}	
\usepackage{amssymb}	
\usepackage{natbib}
\usepackage{times}
\usepackage{multirow}

\usepackage{subfigure}
\subfiguretopcaptrue
\usepackage{url}

\usepackage{etoolbox}

\shortauthors{Zhang \& Yan}

\begin{document}
\title{Polarization of fluorescence lines: tracing magnetic field from circumstellar medium to early universe}

\author[0000-0003-2840-6152]{Heshou Zhang}
\affiliation{Deutsches Elektronen-Synchrotron DESY, Platanenallee 6, D-15738 Zeuthen, Germany}
\affiliation{Institut f$\ddot{u}$r Physik und Astronomie, Universit$\ddot{a}$t Potsdam, Haus 28, Karl-Liebknecht-Str. 24/25, D-14476 Potsdam, Germany}

\author[0000-0003-2560-8066]{Huirong Yan}\thanks{huirong.yan@desy.de}
\affiliation{Deutsches Elektronen-Synchrotron DESY, Platanenallee 6, D-15738 Zeuthen, Germany}
\affiliation{Institut f$\ddot{u}$r Physik und Astronomie, Universit$\ddot{a}$t Potsdam, Haus 28, Karl-Liebknecht-Str. 24/25, D-14476 Potsdam, Germany}

\begin{abstract}
Fluorescence emission lines are broadly applied in observation for diffuse medium in the universe. They are normally observed around strong pumping source, tracing the gas in circumstellar medium, reflection nebula, and H\,{\sc ii} regions, etc. They reside in UV/optical and infrared bands and hence could be directly observed with ground-base telescopes.
In this letter, we demonstrate the polarization of fluorescence lines as a magnetic field tracer arising from ground state atomic alignment in diffuse medium, including our solar system, supernova remnants (SNRs), as well as quasi-stellar object (QSO) host galaxies.
Two types of fluorescence emissions are considered: the primary fluorescence from the excited states; and the secondary fluorescence from the metastable state (forbidden lines).
We find that the synergy of these lines could measure three-dimensional magnetic direction: the polarizations of the primary fluorescence lines could reveal the magnetic polar angle along the line-of-sight, whereas the polarization of forbidden lines traces the plane-of-sky magnetic direction. The expected degree of polarization is $P>10\%$. Polarizations of both types of fluorescence emissions have shown strong potential for observations, and are applicable to measure magnetic field within and beyond our galaxy.\\
\end{abstract}

\section{Introduction}

Atomic/ionic lines play important roles in tracing the chemical evolution and gas structure in the Universe. The optical emission lines from elements such as Sulfur, Oxygen, Carbon, and even heavier Iron, Chromium, have been broadly utilized in astronomy. For decades, the emission lines have been used to study the universe near and far: comet and satellite in our solar system \citep{O1egLocal1,O1egIo}, circumstellar regions in Magellanic Cloud \citep{O1egSMC1,O1egLMC1}, chemistry of supernova remnant \citep{O1egNebular,O1egSNR1}, and even other galaxies \citep{O1egExtra1,O1egExtra2}, etc.

\begin{figure*}
\centering
\includegraphics[width=0.98\textwidth, height=0.32\textheight]{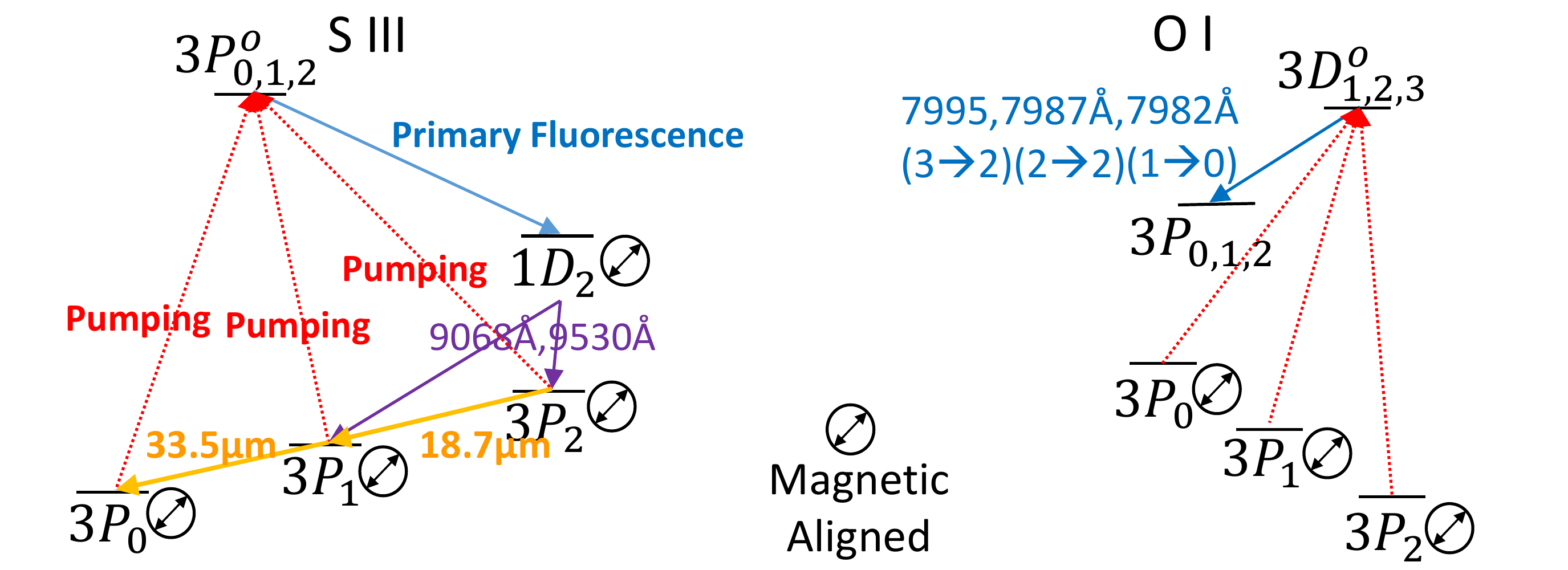}
\caption{ (a) The schematics for atomic transitions of fluorescence emissions. S\,{\sc iii} and O\,{\sc i} are presented as examples. The ground state and meta-stable state are magnetically aligned. The pumping is marked with red dotted lines. The primary fluorescence lines are marked with solid blue arrows. The secondary fluorescence lines are marked in purple arrows. The upper and lower energy levels for lines of our interest are marked on the corresponding transitions. As a result of GSA, the transitions between sublevels on the ground state (S\,{\sc iii}$\lambda\lambda18.7, 33.5\mu{m}$) and between the metastable level and the ground state (S\,{\sc iii}$\lambda\lambda9068, 9530\AA$) are magnetically aligned. Although the excited states are not magnetically aligned, the primary fluorescence lines (e.g., O\,{\sc i}$\lambda\lambda7982$\AA(1-0),$7987$\AA(2-2), and ,$7995$\AA(3-2)) are GSA influenced due to pumping.
}\label{Fig:transition}
\end{figure*}

In the presence of the strong radiation source, the emission lines in optical and near-infrared bands are mostly from fluorescence. The UV photons from the source  pump the atoms from the ground state to the excited states, and then to a lower energy state through the fluorescence emission. The first fluorescence photon from the excited states to a lower level is {\it primary fluorescence}, and further emission from the lower level is {\it secondary fluorescence}, etc (see transition examples demonstrated in Fig.~\ref{Fig:transition}). It has been proposed that polarization of atomic fluorescence lines traces magnetic fields in diffuse media \citep{YLhanle,Nordsieck08}. The emission line polarization is modulated by magnetic field through ground state atomic alignment (GSA, \citealp{YLhyf,YL12,SY13,ZYD15}), whereas the forbidden lines are polarized due to the alignment on the metastable levels \citep{YLhanle, ZY18}. Recent observational study \cite{Zhang20ApJL} has shown that three-dimensional (3D) sub-Gauss magnetic field can be observed with the linear polarization from atomic absorption lines resulting from GSA as predicted before \citep{YLfine}.

\begin{figure*}
\centering
\includegraphics[width=0.9\textwidth, height=0.45\textheight]{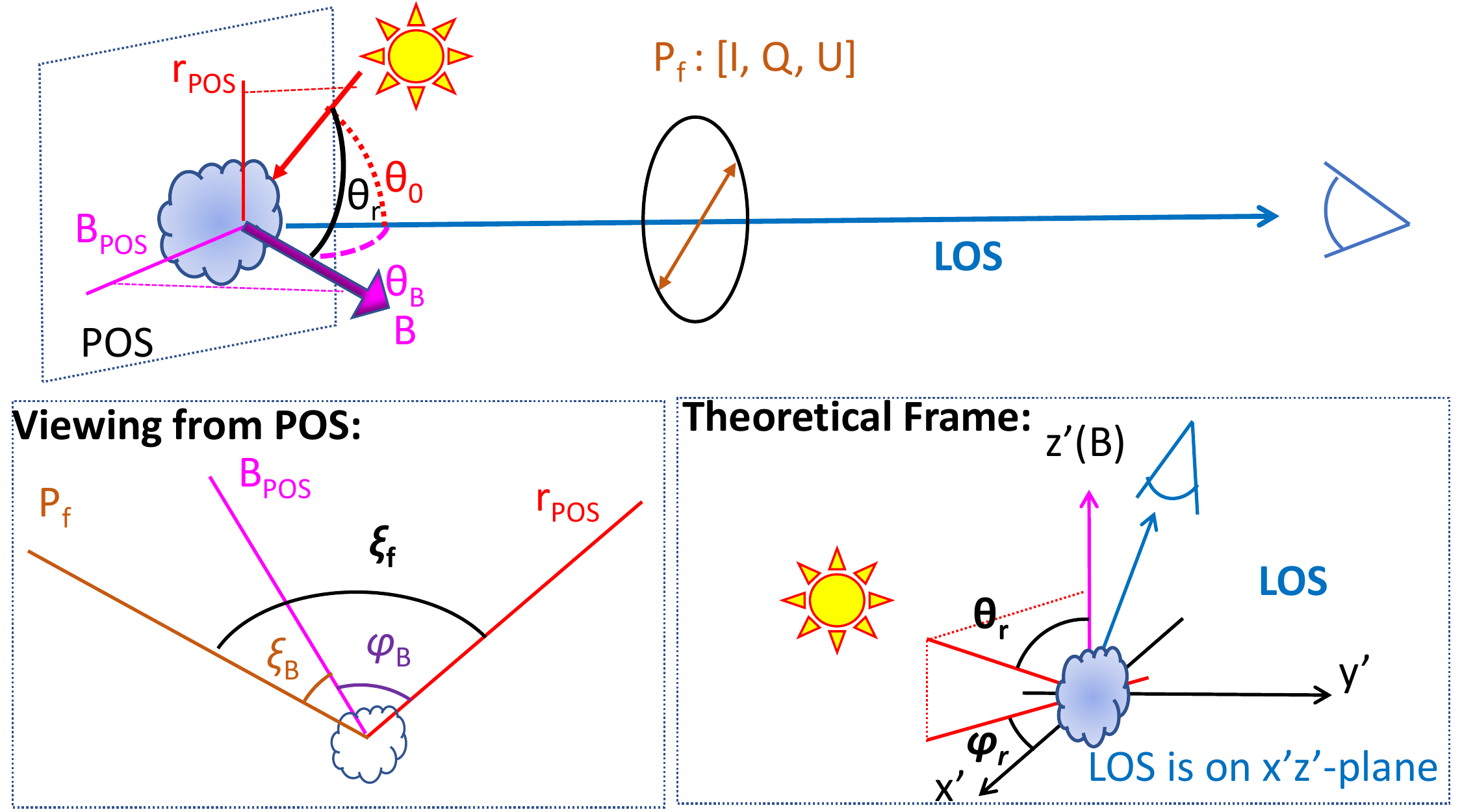}
\caption{The geometry for GSA observation. {\it Top:} The alignment happens on atoms by the vicinity pumping source. The polarized fluorescence lines is $P_f$. The scattering angle is $\theta_0$. The magnetic and radiation directions are projected on the plane-of-sky (POS) as $B_{POS}$ and $r_{POS}$, respectively. $\theta_r$ is the angle between radiation and magnetic field. $\theta_B$ and $\varphi_B$ are the polar and azimuth angles for magnetic field in the observation frame.
{\it Bottom left:} what we can observe on the POS. The angle between the polarization $P_f$ and the radiation $r_{POS}$ is $\xi_f$.
{\it Bottom right:} The theoretical reference frame. $Oz'-$ is the magnetic field direction and the line-of-sight (LOS) is on the $x'Oz'-$plane. $\theta_r$ and $\varphi_r$ is the polar and azimuth angle in this frame.
}\label{Fig:geometry}
\end{figure*}

GSA is applicable to the diffuse medium where the magnetic field strength is normally at sub-Gauss scale. In diffuse interstellar medium (ISM), UV/optical photons  excite atoms from the ground state to the excited states due to optical pumping. The pumping source can be stars, clusters, or even QSOs (quasi-stellar objects) that provide anisotropic radiation field. The angular momenta of the atoms are aligned according to the anisotropic pumping. In diffuse ISM, the Larmor precession rate is faster than the photo-excitation rate from the ground state ($\nu_{L}\ll \tau^{-1}_R$, where $\tau^{-1}_R\equiv BI^{\ast}$, $B$ is the Einstein coefficient for absorption and $I^{\ast}$ is the radiation intensity). Hence, the atoms on the ground state are magnetically aligned. In GSA regime, the spontaneous emission rates from the excited states are faster than the magnetic precession rate ($A>\nu_{L}$, $A$ is the Einstein coefficient for emission). The geometry of this scattering process is shown in Fig.~\ref{Fig:geometry}.

Although the magnetic precession is negligible on the excited states for diffuse medium with subGauss magnetic field, the magnetic alignment on the ground state  modulate the excited states through pumping. As a result, the polarization of the primary fluorescence lines carries magnetic field information in GSA regime. The obvious challenge here is how to deduce the magnetic field information from the primary fluorescence. Additionally, the magnetic alignment is not limited to the ground state, but also the meta-stable level responsible for secondary fluorescence emission. As an example in diffuse ISM, a $6{\mu}G$ magnetic field  yield Larmor precession rate $\nu_{L}\simeq10^2 s^{-1}$. The spontaneous emission rate from metastable state of atoms is typically between $10^{-2}s^{-1}\sim10^{-5}s^{-1}$). Thus, the magnetic precession rate dominates over the escape rate on the metastable state, and the polarization of emission from this state is magnetically aligned.

In this paper, we perform comprehensive calculations for these two types of fluorescence lines. We find an intriguing relation between the primary fluorescence polarization and magnetic polar angle along the line-of-sight. The secondary fluorescence from the metastable level can produce strong polarization signals in different astronomical environments. We  demonstrate that the polarization of fluorescence lines is a powerful magnetic tracer in diffuse medium within and beyond our galaxy.

\section{Physics and formalisms}
The transition equations for density matrix at excited states, metastable levels and the ground state without collision were shown in Eqs.(1-3,29-30) in \cite{YLhanle}. In this study, we take into account the collision effect. There are three types of collisions: superelastic $C_S$, inelastic $C_I$, elastic collisions $C_E$ (see definition in \citealt{Landi04}). Accounting for all the collision terms, the transition equations are modulated into:
\begin{equation}
\label{upperevolcol}
\begin{split}
\dot{\rho}_q^k(J_u)+2\pi i\nu_L g_u q\rho_q^k(J_u)&=-\sum_{\substack{J_l}}A_{ul}\rho_q^k(J_u)\\
&+\sum_{\substack{J_{l}k'}}[J_l]\left[\delta_{q0}\delta_{kk'}p_{k'}(J_u,J_l)B_{lu}\bar{J}^0_0
+\sum_{\substack{Qq'}}r_{kk'}(J_u,J_l,Q,q')B_{lu}\bar{J}^2_Q\right]\rho^{k'}_{-q'}(J_l),
\end{split}
\end{equation}
\begin{equation}
\label{metaevolcol}
\begin{split}
&\dot{\rho}_q^k(J_l)+2\pi i\nu_L g_l q\rho_q^k(J_l)=\sum_{\substack{J_u}}p_k(J_u,J_l)[J_u]A_{ul}\rho_q^k(J_u)
-\sum_{\substack{J'_l}}A_m\mid_{E'<E}\rho_q^k(J_l)+\sum_{\substack{J'_l}}p_k(J'_l,J_l)[J'_l]A_m\mid_{E'>E}\rho_q^k(J'_l)\\
&-\sum_{\substack{J_{u}k'}}\left[\delta_{kk'}B_{lu}\bar{J}^0_0+\sum_{\substack{Qq'}}s_{kk'}(J_u,J_l,Q,q')B_{lu}\bar{J}^2_Q\right]\rho^{k'}_{-q'}(J_l)
+\sum_{\substack{J'_l}}\sqrt{\frac{[J'_l]}{[J_l]}}\left(C_S^{(k)}(J_l,J'_l)\mid_{E'>E}+C_I^{(k)}(J_l,J'_l)\mid_{E'<E}\right)\rho_q^k(J'_l)\\
&-\sum_{\substack{J'_l}}\left[C_I^{(0)}(J'_l,J_l)\mid_{E'>E}+C_S^{(0)}\mid_{E'<E}\right]\rho_q^k(J_l)
-D^{(k)}(J_l)\rho_q^k(J_l).
\end{split}
\end{equation}
in which
\begin{equation}
\begin{split}
&D^{(k)}(J_l)=C_E^{(0)}(J_l)-C_E^{(k)}(J_l); D^{(0)}(J_l)=0;\\
&p_k(J_u,J_l)=(-1)^{J_u+J_l+1}\left\{\begin{array}{ccc}J_l&J_l&k\\J_u&J_u&1\end{array}\right\},\\
&r_{kk'}(J_u,J_l,Q,q)=(3[k,k',2])^{\frac{1}{2}}\left\{\begin{array}{ccc}1&J_u&J_l\\1&J_u&J_l\\2&k&k'\end{array}\right\}\left(\begin{array}{ccc}k&k'&K\\q&q'&Q\end{array}\right),\\
&s_{kk'}(J_u,J_l,Q,q)=(-1)^{J_l-J_u+1}[J_l](3[k,k',K])^\frac{1}{2}
\left(\begin{array}{ccc}k&k'&2\\q&q'&Q\end{array}\right) \left\{\begin{array}{ccc}1&1&2\\J_l&J_l&J_u\end{array}\right\}\left\{\begin{array}{ccc}k&k'&2\\J_l&J_l&J_l\end{array}\right\}.
\end{split}
\end{equation}
The geometry is demonstrated in Fig.~\ref{Fig:geometry}. The quantities $J_{u}$ and $J_{l}$ are the total angular momentum quantum numbers for the upper and lower levels, respectively. The quantities $\rho_q^k$ and $\bar{J}_Q^K$ are irreducible density matrices for the atoms and the incident radiation, respectively. $6-j$ and $9-j$ symbols are represented by the matrices with $"\{$ $\}"$, whereas $3-j$ symbols are indicated by the matrices with $"()"$. {\em The second terms on the left side of the equations stand for the magnetic realignment.} The terms on the right side represent spontaneous emissions, the excitations from lower levels, the magnetic dipole transitions among ground and metastable states, and the three types of collisions, respectively. Note that the spontaneous emission, collision, and magnetic realignment conserve $k$ and $q$.

The superelastic collision and inelastic collision follow the Einstein-Milne relation $C_S^{(0)}(J_l,J_u)=\frac{[J_l]}{[J_u]}e^{\frac{E(J_u)-E(J_l)}{k_BT_c}}C_I^{(0)}(J_u,J_l)$.
The collision de-excitation rate due to electron and hydrogen can be expressed \citep[see][]{Osterbrock06book}:
\begin{equation}\label{coleq}
C_{ul,e}=n_e\frac{8.629\times10^{-6}\Upsilon_{lu}}{T^{1/2}[J_u]},
C_{ul,H}\simeq n_H\frac{1.1\times10^{-10}\Upsilon_{lu}}{T^{1/2}[J_u]}
\end{equation}
where $I$ is the ionization potential of the atom and $E(J)$ is the energy level in $eV$. $\mu$ is the atomic mass.

The inelastic and superinelastic collision rate between upper and lower levels follow \citep[see][]{Landi04}:
\begin{equation}\label{ulcolrate}
\begin{split}
&C_I^{(k)}(J_u,J_l)=(-1)^{k}\frac{\left\{\begin{array}{ccc}J_u&J_u&k\\J_l&J_l&\tilde{K}\end{array}\right\}}{\left\{\begin{array}{ccc}J_u&J_u&0\\J_l&J_l&\tilde{K}\end{array}\right\}}C_I^{(0)}(J_u,J_l),
C_S^{(k)}(J_l,J_u)=(-1)^{k}\frac{\left\{\begin{array}{ccc}J_l&J_l&k\\J_u&J_u&\tilde{K}\end{array}\right\}}{\left\{\begin{array}{ccc}J_l&J_l&0\\J_u&J_u&\tilde{K}\end{array}\right\}}C_S^{(0)}(J_l,J_u)\\
&C_{lu}=C_I^{(0)}(J_u,J_l), C_{ul}=C_S^{(0)}(J_l,J_u)
\end{split}
\end{equation}
The depolarizing rate due to elastic collision can be approximated by \citep{Lamb71,Landi04}:
\begin{equation}\label{coleqdk}
D^{(k)}\simeq 2\times10^{-10}\left(\frac{13.6}{I-E(J)}\right)^{0.8}\left[T\left(1+\frac{1}{\mu}\right)\right]^{0.3} n_H
\end{equation}

To solve Eq.~\eqref{upperevolcol}, magnetic realignment on the levels in the excited states can be neglected because the spontaneous emission rate is much higher than magnetic precession rate in diffuse medium ($\nu_L\ll A_{ul}$). On the other hand, the magnetic precession rate is much higher than the photon excitation rate of the atoms on the ground state and metastable levels ($\nu_L\gg B_{lu}I^\ast$). Thus, $q=0$ for the magnetic aligned sublevels in Eq.~\eqref{metaevolcol}.

The density matrices for the excited states can be obtained from Eq.~\eqref{upperevolcol}:
\begin{equation}
\label{upperdens}
\begin{split}
\rho^0_0(J_u)&=\bar{J}^0_0\sum_{\substack{l}}\frac{B_{lu}[J_l]}{\sum_{\substack{J_l}}A_{ul}}p_0(J_u,J_l)\rho^0_0(J_l)+\bar{J}^2_0\sum_{\substack{l}}\frac{B_{lu}[J_l]}{\sum_{\substack{J_l}}A_{ul}}r_{02}(J_u,J_l,0,0)\rho^2_0(J_l)\\
\rho^2_0(J_u)&=\bar{J}^0_0\sum_{\substack{l}}\frac{B_{lu}[J_l]}{\sum_{\substack{J_l}}A_{ul}}p_2(J_u,J_l)\rho^2_0(J_l)+\bar{J}^2_0\sum_{\substack{l}}\frac{B_{lu}[J_l]}{\sum_{\substack{J_l}}A_{ul}}(r_{20}(J_u,J_l,0,0)\rho^0_0(J_l)\\
&+r_{22}(J_u,J_l,0,0)\rho^2_0(J_l)+r_{24}(J_u,J_l,0,0)\rho^4_0(J_l))\\
\rho^2_{\pm1}(J_u)&=\bar{J}^2_{\mp1}\sum_{\substack{l}}\frac{B_{lu}[J_l]}{\sum_{\substack{J_l}}A_{ul}}(r_{20}(J_u,J_l,-1,1)\rho^0_0(J_l)+r_{22}(J_u,J_l,-1,1)\rho^2_0(J_l)+r_{24}(J_u,J_l,-1,1)\rho^4_0(J_l))\\
\rho^2_{\pm2}(J_u)&=\bar{J}^2_{\mp2}\sum_{\substack{l}}\frac{B_{lu}[J_l]}{\sum_{\substack{J_l}}A_{ul}}(r_{20}(J_u,J_l,-2,2)\rho^0_0(J_l)+r_{22}(J_u,J_l,-2,2)\rho^2_0(J_l)+r_{24}(J_u,J_l,-2,2)\rho^2_0(J_l))
\end{split}
\end{equation}

\begin{figure*}
\label{Fig:fluoRegime}
\centering
\includegraphics[width=0.98\textwidth]{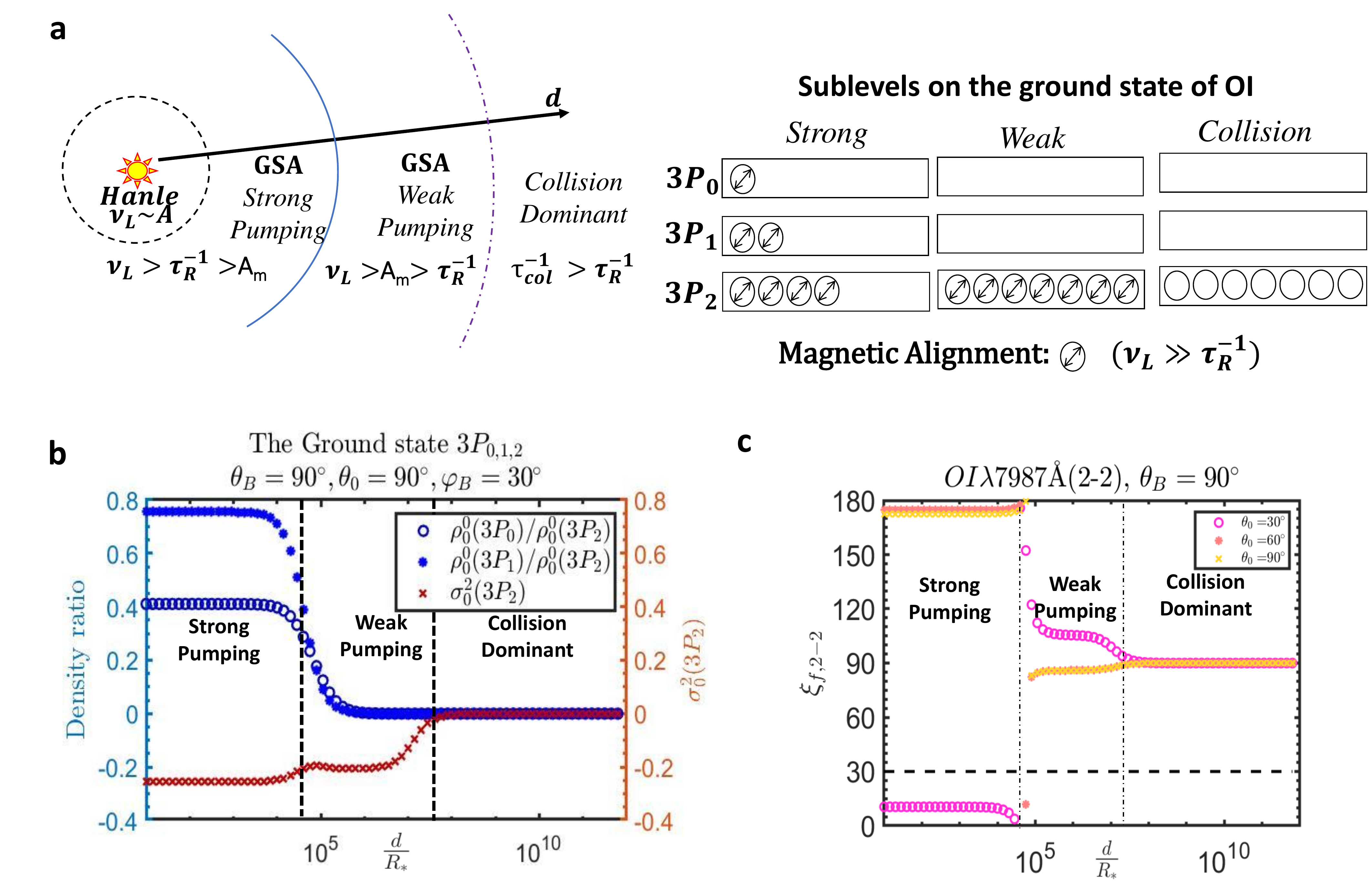}
\caption
{(a) {\it Left:} Different pumping regimes are defined based on the comparison between pumping rate ($\tau^{-1}_R$), magnetic precession rate ($\nu_L$), spontaneous emission rate from the metastable level $A_m$, and collisional rate ($\tau^{-1}_R$). {\it Right:} Distribution of atoms on the sublevels of the ground state in different pumping regimes. The circle with double arrows are magnetically aligned levels and atoms in collision dominant regime is not aligned.(b) The ground state density ratios among different sub-levels (blue points, left $y-$axis) and the ground level alignment parameter ($\sigma^2_0(3P_2)$, red points, right $y-$axis). $x-$axis is the ratio between the distance of the medium to the source and the pumping star radius. The results are separated into three different regimes in accordance with the decreasing optical pumping rates as marked in (a).
(c) The polarization angles of O\,{\sc i}$\lambda7987$\AA(2-2) in different regimes for different scattering angles ($\theta_0=30^\circ, 60^\circ, 90^\circ$).}
\end{figure*}

\section{3D Magnetic field measurement with fluorescence lines}

We use the fluorescence from O\,{\sc i} and S\,{\sc iii} as examples to illustrate our calculations. We  take three lines from the neutral Oxygen {\it primary fluorescence} multiplets as examples: O\,{\sc i}$\lambda\lambda7982$\AA(1-0),$7987$\AA(2-2), and ,$7995$\AA(3-2) (see schematics in Fig.~\ref{Fig:transition}). The upper level of the multiplets is $3D^\circ_{J=1,2,3}$($12.539eV$), directly excited from the ground state, whereas the lower level is $3P_{J=0,1,2}$($10.989eV$).
We  use S\,{\sc iii} for secondary fluorescence from the metastable state. For the doubly-ionized Sulfur doublet S\,{\sc iii}$\lambda\lambda9068.6,9530.6$\AA, the upper level is the magnetically aligned meta-stable state $1D_{J_m=2}$($1.404eV$). These metastable levels are populated due to the spontaneous emission from the excited states  (see schematics in  Fig.~\ref{Fig:transition}). The Einstein coefficient for emission from metastable state is relatively low comparing to the permitted emission. Hence, we take into account the collision processes in our calculations.
The comprehensive atomic transition equations are in Eq.~\eqref{upperevolcol}. We consider the aligned atoms are in warm ionized medium (WIM) with temperature 8000K, hydrogen density $n_H=0.25 cm^{-3}$, ionization fraction $\sim0.7$ \citep{WIM77}. The inelastic and superelastic collisions are thereby dominated by the electrons. Based on the calculations presented in the previous section, the collisional excitation rates for neutral Oxygen and doubly-ionized Sulfur in WIM are listed table~\ref{tab:col}.

\begin{table*}[ht]
\caption{\label{tab:col} The collision excitation rates (unit $s^{-1}$) in the warm ionized medium (WIM, $T_{eff}=8000K$, hydrogen density $n_H=0.25 cm^{-3}$, ionization fraction $\sim0.7$ \citealt{WIM77}). Different rows in the table are for the level of the atom before collision, and different columns in the table are for that after the collision, i.e., $E_{from}<E_{to}$  for inelastic collision $C_I$, $E_{from}>E_{to}$ for superelastic collision $C_S$, and $E_{from}=E_{to}$ for elastic collision $D$. Collision strengths are calculated according to \citealt{O1col1,S3col,DraineBook}.}
\begin{center}
\begin{tabular}{cccccc}
\hline\hline
\multicolumn{2}{c}{\multirow{2}{*}{S\,{\sc iii}}} & \multicolumn{4}{c}{$E_{to}$} \\
\multicolumn{2}{c}{{ }} & $0eV (3P_0)$ & $0.037eV (3P_1)$ & $0.103eV (3P_2)$ & $1.404eV (1D_2)$ \\
\hline
\multirow{4}{*}{$E_{from}$} & $0eV (3P_0)$	&	3.52E-10	&	6.67E-08	&	1.94E-08	&	1.71E-09	\\
& $0.037eV (3P_1)$	&	2.34E-08	&	3.53E-10	&	3.95E-08	&	1.71E-09	\\
& $0.103eV (3P_2)$	&	4.50E-09	&	2.76E-08	&	3.53E-10	&	1.71E-09	\\
& $1.404eV (1D_2)$	&	2.63E-09	&	7.88E-09	&	1.31E-08	&	3.64E-10	\\
\hline
\multicolumn{2}{c}{\multirow{2}{*}{O\,{\sc i}}} & \multicolumn{4}{c}{$E_{to}$} \\
\multicolumn{2}{c}{{ }} & $0eV (3P_2)$ & $0.019eV (3P_1)$ & $0.028eV (3P_0)$ & $1.967eV (1D_2)$ \\
\hline
\multirow{4}{*}{$E_{from}$} & $0eV (3P_2)$	&	7.55E-10	&	2.58E-10	&	8.17E-11	&	2.28E-11	\\
& $0.019eV (3P_1)$	&	4.43E-10	&	7.56E-10	&	1.08E-10	&	2.28E-11	\\
& $0.028eV (3P_0)$	&	4.26E-10	&	3.37E-10	&	7.56E-10	&	2.28E-11	\\
& $1.967eV (1D_2)$	&	3.97E-10	&	2.38E-10	&	7.93E-11	&	8.55E-10	\\
\hline\hline
\end{tabular}
\end{center}
\end{table*}

\begin{figure}
\includegraphics[width=0.98\columnwidth]{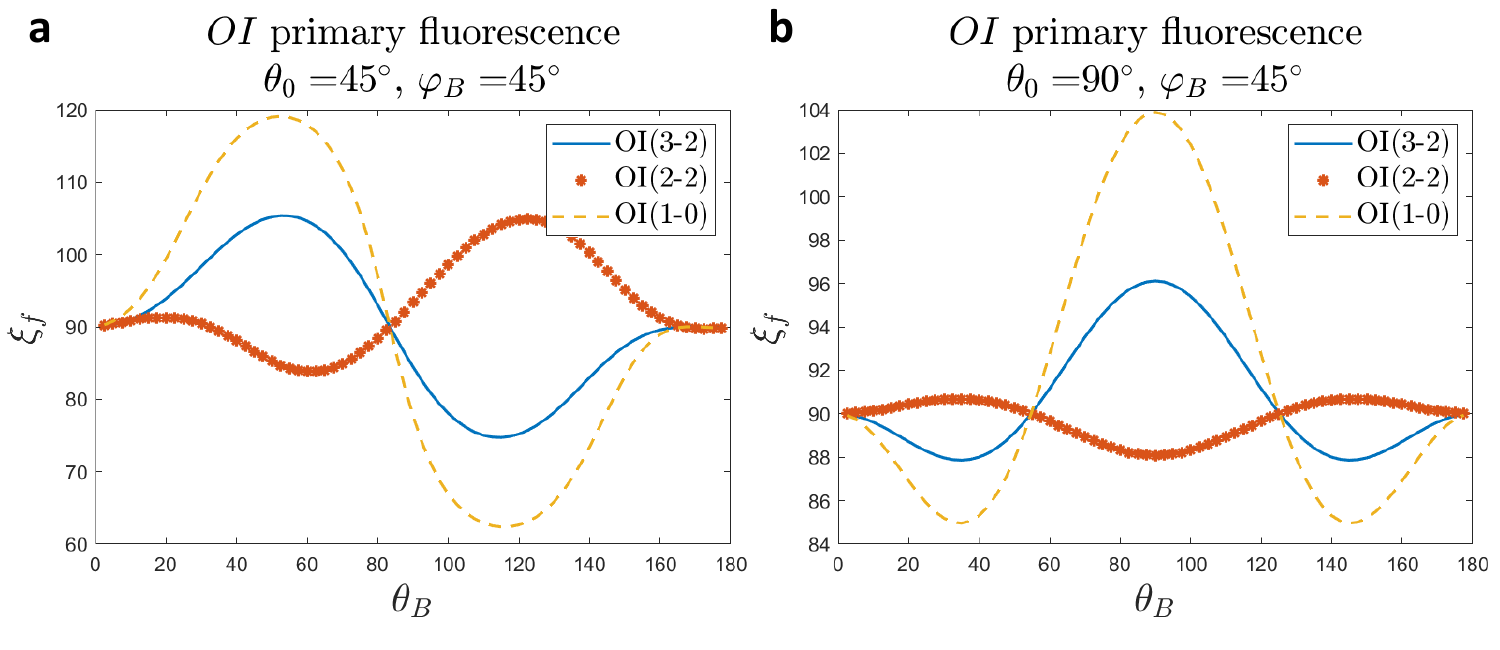}
\caption{The polarization directions for 3 primary fluorescence lines (O\,{\sc i}$\lambda\lambda7982$\AA(1-0),$7987$\AA(2-2), and ,$7995$\AA(3-2)). The geometry is $\varphi_B=45^\circ$, (a) $\theta_0=45^\circ$,  (b) $\theta_0=90^\circ$. $x-$axis is $\theta_B$. $y-$axis is the observed polarization angle.
}\label{Fig:Linescan}
\end{figure}

\subsection{Polarization of Primary fluorescence}
By solving the transition equations Eq.~\eqref{upperevolcol}, the density matrix ($\rho^k_q$) and the alignment parameters ($\sigma^2_0\equiv\rho^2_0/\rho^0_0$) under the influence of magnetic alignment are obtained. Let us consider diffuse medium in the vicinity of a blue giant as an example (effective radius $0.1AU$, effective temperature $T_{eff}=16000K$). We compare the results at different distance from pumping source, indicating the change from pumping dominant (near, $\tau_R^{-1}>\tau_{col}^{-1}$) to collision dominant (far, $\tau_R^{-1}<\tau_{col}^{-1}$) regime. The density ratio between the upper sublevels and the ground level on the ground state ($\frac{\rho^0_0(3P_{0,1})}{\rho^0_0(3P_{2})}$) are calculated in Fig.~\ref{Fig:fluoRegime}b. The alignment on the ground state $\sigma^2_0(3D_2)$ is also calculated with respect to the distance from the pumping source. Fig.~\ref{Fig:fluoRegime} shows that the medium is categorized into four zones: (1) Hanle regime where magnetic precession rate is comparable to the spontaneous emission rate from the excited states ($\tau_R^{-1}\sim A$, which only applies at the stellar surface with magnetic field larger than Gauss scale); (2) GSA strong pumping (pumping from the ground state faster than spontaneous emission rate from the metastable levels, $\tau_R^{-1}>A_m>\tau_{col}^{-1}$). In current calculation, the medium with up to a few thousand AUs of the distance from the central star is suitable for this regime; (3) GSA weak pumping (pumping from the ground state slower than spontaneous emission rate from the metastable levels but faster than the collision rate, $A_m>\tau_R^{-1}>\tau_{col}^{-1}$). In current calculation, the medium with up to several pcs of the distance from the central star is suitable for this regime; and (4) collision dominant (pumping from the ground state slower than the collision rate, $A_m>\tau_{col}^{-1}>\tau_R^{-1}$). $\sigma^2_0(3D_2)$ calculation in Fig.~\ref{Fig:fluoRegime}b demonstrates that the alignment on the ground state happens in the first two regimes but not in the collision dominant regime. Based on the density distribution among different sub-levels on the ground state, the alignment from all the metastable levels are not negligible in the strong pumping regime, whereas only the alignment on the ground level contribution is dominant in the weak pumping regimes. We illustrate the ground state alignment distribution among different sublevels in Fig.~\ref{Fig:fluoRegime}a.

We investigate the primary fluorescence by first considering the upper level density on the excited state. The densities on the excited states ($\rho^k_q(J_u)$) can be expressed by the ground level density matrix (see Eq.~\ref{upperdens}). Hence, the polarization for the primary fluorescence can be expressed by the density matrix on the excited states:
\begin{equation}\label{pemprimary}
\begin{split}
&\frac{Q_f}{I}=\frac{w^2_{J_uJ_l}\sum_{\substack{Q}}(\textit{J}^2_Q(1)\cos2\varphi_B+\textit{J}^2_Q(2)\sin2\varphi_B)\sigma^2_Q(3D^\circ_2)}{1+w^2_{J_uJ_l}\sum_{\substack{Q}}\textit{J}^2_Q(0)\sigma^2_Q(3D^\circ_2)},\\
&\frac{U_f}{I}=\frac{w^2_{J_uJ_l}\sum_{\substack{Q}}(-\textit{J}^2_Q(1)\sin2\varphi_B+\sum_{\substack{Q}}\textit{J}^2_Q(2)\cos2\varphi_B)\sigma^2_Q(3D^\circ_2)}{1+w^2_{J_uJ_l}\sum_{\substack{Q}}\textit{J}^2_Q(0)\sigma^2_Q(3D^\circ_2)},\\
&w^2_{2,2}=-0.5916; w^2_{1,0}=1; Q=0,\pm1, \pm2.
\end{split}
\end{equation}
The polarization angle expected for the primary fluorescence is $\xi_f=\frac{1}{2}\arctan\frac{U_f(J_u-J_l)}{Q_f(J_u-J_l)}$.

We first consider the scenario without GSA effect, i.e., the ground state of the atoms are not aligned ($\rho^k_q(J_l)=0$, for $k>0$). In this regime, we insert Eq.~\ref{upperdens} into Eq.~\ref{pemprimary} and obtain $\xi_f\equiv 90^\circ$. Hence, without the influence of magnetic field, the polarization for different fluorescence lines is scattered perpendicular to the radiation direction.

Fig.~\ref{Fig:fluoRegime}c demonstrates the primary fluorescence line O\,{\sc i}$\lambda7987$\AA(2-2) with different scattering angles ($\theta_0=30^\circ, 60^\circ, 90^\circ$). We consider here the magnetic direction on the plane of sky with a $30^\circ-$angle difference comparing to the radiation projection ($\theta_B=90^\circ, \varphi_B=30^\circ$). In the collision dominant regime, there is no alignment on the ground state. Hence, the directions of polarization follow the scattering direction, perpendicular to the anisotropic radiation direction. However, the higher order density matrices on the ground state are non-zero in two GSA pumping regimes ($\rho^{2}_0(J_l)\neq0$, see Fig.~\ref{Fig:fluoRegime}b). Hence, the density matrices on the excited states are modulated accordingly. The polarization of the primary fluorescence is different for strong and weak pumping regime.

\subsection{Magnetic Polar angle}

The polarization directions of different primary fluorescence lines are investigated under the influence of magnetic alignment on the ground state.
We note that, the POS magnetic field direction can be acquired based on the fine-structure lines, which are magnetically aligned and their polarization directions are the magnetic POS direction with a $90^\circ-$degeneracy. Fig.~\ref{Fig:Linescan} compares the polarization directions of three primary fluorescence lines, O\,{\sc i}$\lambda\lambda7982$\AA(1-0),$7987$\AA(2-2), and ,$7995$\AA(3-2). As can be seen from Fig.~\ref{Fig:Linescan}, the magnetic alignment on the ground state results in different polarization directions for these lines and their angle differences are changing with the magnetic polar angle $\theta_B$.

\begin{figure*}
\includegraphics[width=0.98\textwidth]{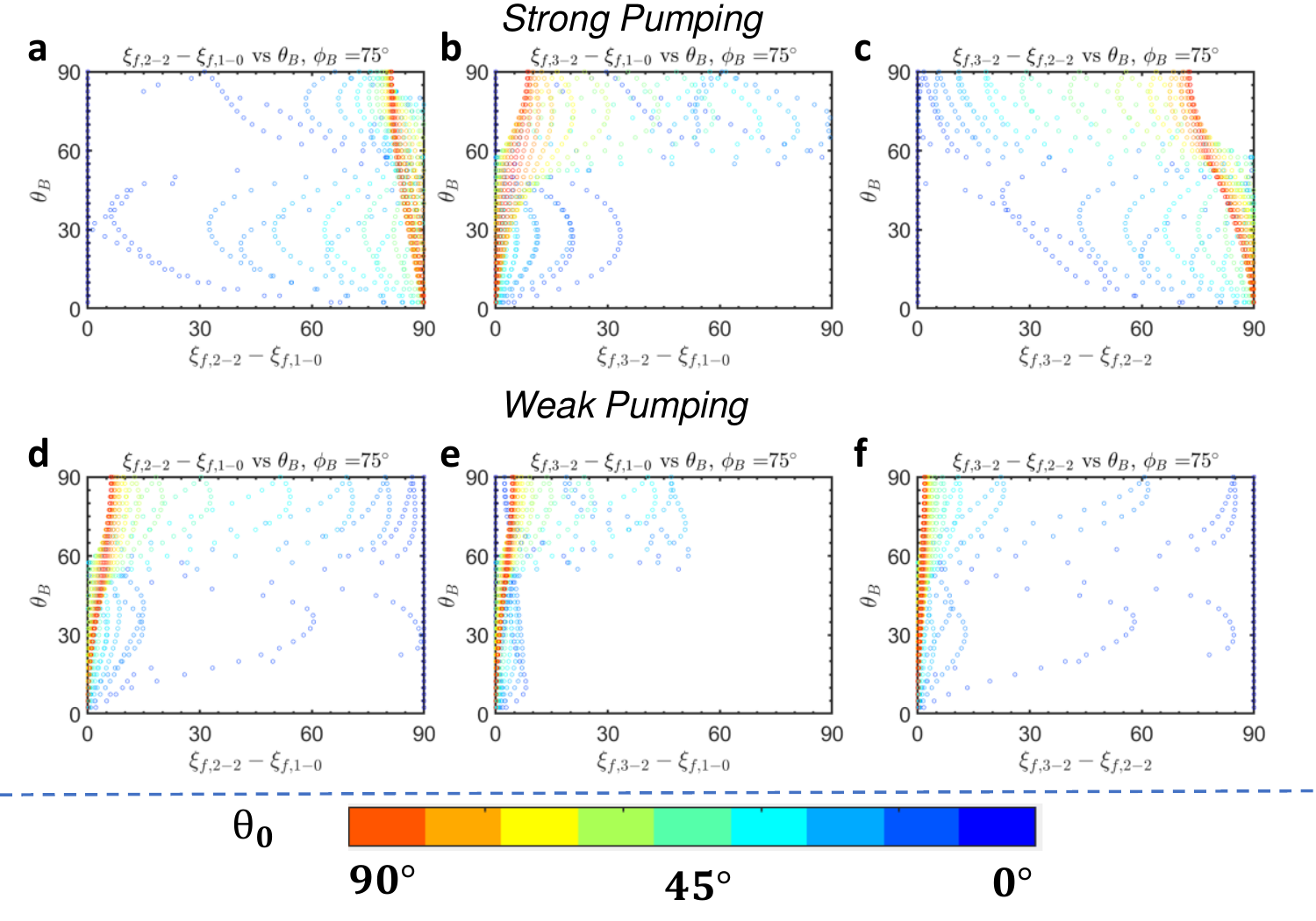}
\caption{The angle difference between the polarization directions of three primary fluorescence lines ((O\,{\sc i}$\lambda\lambda7982$\AA(1-0),$7987$\AA(2-2), and ,$7995$\AA(3-2)) at (a-c) strong pumping; (d-f) weak pumping regimes. $x-$axis is the angle between (a, d) $\xi_{1-0}$ and $\xi_{2-2}$; (b, e) $\xi_{1-0}$ and $\xi_{3-2}$; (c, f)$\xi_{2-2}$ and $\xi_{3-2}$. $y-$axis is $\theta_B$. Different scattering angles ($\theta_0$) are presented with different color. The POS magnetic direction is at $\theta_B=75^\circ$.
}\label{Fig:Tscan}
\end{figure*}

Furthermore, we perform a more comprehensive investigation on the relation between the polarization angle differences of these primary fluorescence lines against the magnetic polar angle from LOS. The results are demonstrated in Fig.~\ref{Fig:Tscan}. Both strong pumping and weak pumping regimes are investigated\footnote{We note that the strong pumping and weak pumping regimes can be distinguished in observation by comparing the pumping rate $\tau^{-1}_R\equiv BI^\ast$ and the fine-structure transition rate on the ground state $A_m$, i.e., $BI^\ast>A_m$ is for strong pumping and $BI^\ast<A_m$ for weak pumping}. Fig.~\ref{Fig:Tscan} shows that by observing the angle between different primary fluorescence polarization, the magnetic polar angle from LOS can be constrained. If polarizations of several primary fluorescence lines are observed, the $\theta_B$ can be obtained with higher precision. We note that the magnetic polar angle is a unique information that is only available with GSA for diffuse ISM.

\subsection{Secondary fluorescence from metastable level}

For the secondary fluorescence, we only focus on the transitions from the metastable level. In GSA regime, the magnetic realignment  dominate over the escape rate from this level ($\nu_{L}>A_m$). Hence, the upper state of these transitions is magnetically aligned. The polarization directions of those lines are either parallel or perpendicular to the POS magnetic field projection. {\it The synergy of polarization direction from primary and secondary fluorescence could measure the 3D magnetic field directly.}

We use S\,{\sc iii}$\lambda\lambda9068.6,9530.6$\AA doublets to study further the degree of polarization in different environments. The polarization of the [S\,{\sc iii}] doublet are:
\begin{equation}\label{pemeq}
\begin{split}
\frac{Q}{I}(9068.6\mbox{\AA})&=\frac{-2.51\sigma^2_0(1.404eV)\sin^2\theta_B\cos2\varphi_B}{4+0.8367\sigma^2_0(1.404eV)(2-3\sin^2\theta_B)},\\
\frac{U}{I}(9068.6\mbox{\AA})&=\frac{-2.51\sigma^2_0(1.404eV)\sin^2\theta_B\sin2\varphi_B}{4+0.8367\sigma^2_0(1.404eV)(2-3\sin^2\theta_B)},\\
\frac{Q}{I}(9530.6\mbox{\AA})&=\frac{2.51\sigma^2_0(1.404eV)\sin^2\theta_B\cos2\varphi_B}{4-0.8367\sigma^2_0(1.404eV)(2-3\sin^2\theta_B)},\\
\frac{U}{I}(9530.6\mbox{\AA})&=\frac{2.51\sigma^2_0(1.404eV)\sin^2\theta_B\sin2\varphi_B}{4-0.8367\sigma^2_0(1.404eV)(2-3\sin^2\theta_B)},.
\end{split}
\end{equation}

\begin{figure}
\includegraphics[width=0.49\columnwidth]{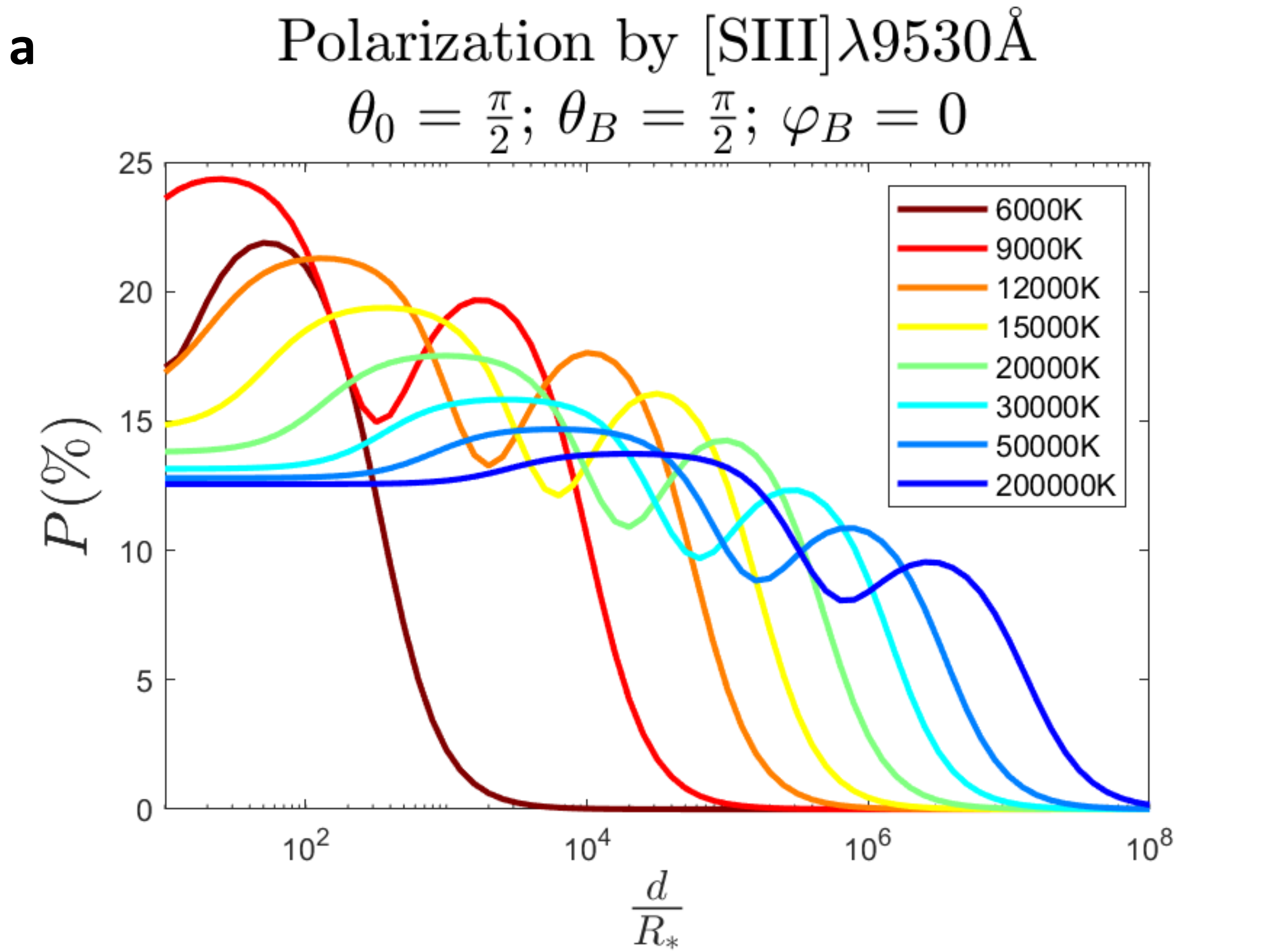}
\includegraphics[width=0.49\columnwidth]{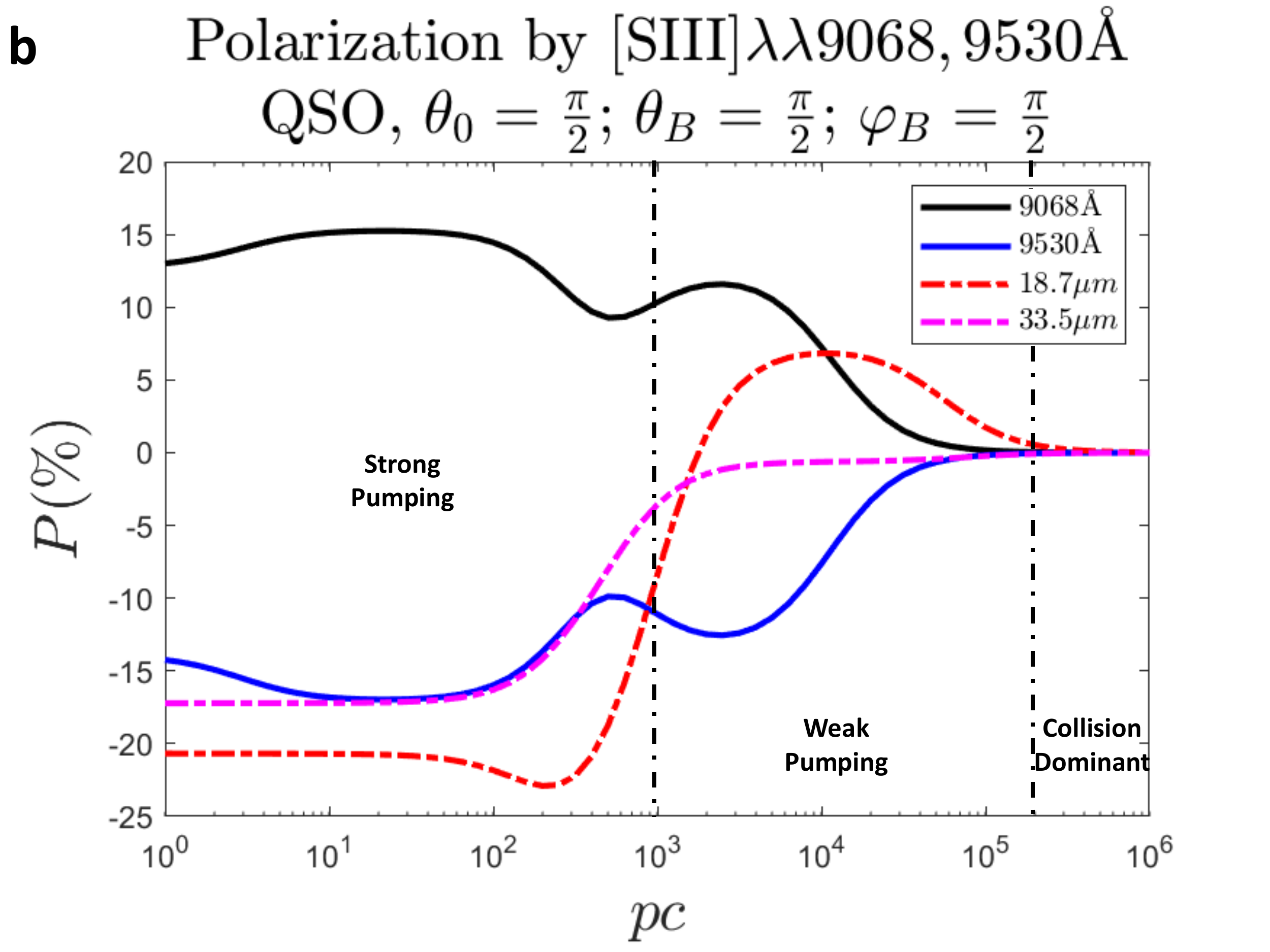}
\caption{(a) The maximum linear polarizations produced for [S\,{\sc iii}]$\lambda9530$\AA\, emission. The geometry is fixed at $\theta_0=90^\circ$; $\theta_B=90^\circ$; $\varphi_B=0^\circ$. The maximum polarization is achieved in this geometry. Different pumping sources are considered, and the effective temperature of the source is marked with different colors. $x-$axis is the ratio between the distance from the source to the medium and the pumping star effective radius $\frac{d}{R_{\ast}}$. The sign of the polarization means the direction of polarization being parallel (positive) or perpendicular (negative) to the magnetic field. (b) The maximum linear polarizations produced with for medium in QSO host Galaxies for doubly-ionized Sulfur lines: secondary fluorescence from metastable levels [S\,{\sc iii}]$\lambda9068$\AA (black), [S\,{\sc iii}]$\lambda9530$\AA (blue); and fine-structure emission within the ground state [S\,{\sc iii}]$18.7\mu{m}$(dash-dotted red), [S\,{\sc iii}]$33.5\mu{m}$(dash-dotted pink). The setup is the same as Fig.\ref{Fig:Tscan}.
}\label{Fig:QSOscan}
\end{figure}

The maximum polarization produced with different pumping stars are compared at the distance ranging from 10 stellar radius to $10^8$ stellar radius ($R_\ast$). We take the geometry when $\theta_0=90^\circ$; $\theta_B=90^\circ$; $\varphi_B=0^\circ$, and the sign of polarization indicates the direction of polarization with respect to magnetic field direction. As shown in Fig.\ref{Fig:QSOscan}a, for all different pumping sources, the maximum polarization can reach more than $10\%$ for [S\,{\sc iii}]$\lambda9530$\AA, in both strong and weak pumping regimes. For all cases, the polarization  decrease to $\sim0$ when the aligned medium is too distant from the pumping source because the collisions become dominant. The polarization direction is parallel to the magnetic field direction. With higher effective temperature star, the maximum potential polarization degree becomes smaller, but the effective distance that can produce the polarization expands. The influence of the collisions is marginal with a higher temperature pumping source.

\subsection{fluorescence lines in QSO host galaxies}

We also calculated the expected polarization of [S\,{\sc iii}]$\lambda\lambda9068,9530$\AA\, doublet from the medium in QSO host galaxies where the quasar provides the UV/Optical photons for pumping. We adopt the QSO emission being a power-law profile $f_\nu\propto\nu^{-0.61}$ when the wavelength $\lambda>912$\AA\, and the luminosity of the QSO continuum $\sim 1000$\AA\, to be $\nu L_\nu\sim10^{47} ergs\cdot s^{-1}$ \citep[see][]{Lusso15}. We consider WIM with the same parameters from the previous subsection. The polarizations at the geometry ($\theta_0=90^\circ$; $\theta_B=90^\circ$; $\varphi_B=0^\circ$) are plotted in Fig.\ref{Fig:QSOscan}. The fine-structure emissions within the ground state [S\,{\sc iii}]$\lambda\lambda18.7,33.5\mu{m}$ are plotted as reference. For [S\,{\sc iii}]$\lambda\lambda9068,9530$\AA\, doublet, the polarization degree is larger than $10\%$ when the medium is within $d\lesssim10kpc$ distance, regardless of strong or weak pumping regimes. The polarization of [S\,{\sc iii}]$\lambda18.7\mu{m}$ is parallel to [S\,{\sc iii}]$\lambda9530$\AA\, in strong pumping regime, whereas parallel to [S\,{\sc iii}]$\lambda9068$\AA\, in weak pumping regime. The synergy of polarizations from different metastable lines in GSA regime could measure the distance from the medium to the pumping source. Additionally, the magnetic field strength can be constrained by comparing the Larmor precession rate and the Einstein coefficients of different lines.

\begin{table}[ht]
\caption{\label{tab:Pmax} Maximum polarization ($\%$) produced in different pumping environments at strong pumping regime ($B_{lu}I^\ast>A_m$). The pumping sources are indicated in the first row. The maximum polarization is calculated by accounting for all different geometries among different distance from the pumping source with strong pumping regime.}
{
\begin{center}
\begin{tabular}{ccccccc}
\hline\hline
\multicolumn{2}{c}{pumping} &	$T_{eff}$(K)&	$T_{eff}$(K)&	$T_{eff}$(K)&	QSO  & The\\
\multicolumn{2}{c}{source} &	$9000$&	$15000$&	$50000$& & Sun\\
\hline
\multirow{2}{*}{C\,{\sc i}}& 9824 ($1D_2\rightarrow3P_1$)&	24.9&	18.5&	4.2&	10.4& 27\\
&9850 ($1D_2\rightarrow3P_2$)&	29.9&	21.1&	4.3&	11.2& 33\\
\hline
\multirow{2}{*}{O\,{\sc i}}& 6300 ($1D_2\rightarrow3P_2$)&	10.3&	11.0&	11.7&	11.6& 8\\
&6363 ($1D_2\rightarrow3P_1$)&	11.1&	11.9&	12.7&	12.5& 8\\
\hline
\multirow{2}{*}{S\,{\sc i}}& 10821 ($1D_2\rightarrow3P_2$)&	29.1&	22.2&	16.3&	17.0& 17\\
&11305 ($1D_2\rightarrow3P_1$)&	24.3&	19.4&	14.7&	15.2& 15\\
\hline
\multirow{2}{*}{S\,{\sc iii}}& 9068 ($1D_2\rightarrow3P_1$)&	18.0&	18.1&	18.1&	18.1& 14\\
&9530 ($1D_2\rightarrow3P_2$)&	16.1&	16.1&	16.2&	16.1& 20\\
\hline\hline
\end{tabular}
\end{center}}
\end{table}

\subsection{Maximum Polarization for secondary fluorescence}

Using similar procedures as we did for fine structure transitions \citep{YLhanle, YL12, ZY18}, we obtained the maximum polarization of the secondary fluorescence lines from metastable levels for different elements. The results for strong pumping regime with different pumping sources are listed in Table~\ref{tab:Pmax}. We also use the sun as pumping source, and consider the medium within the distance of $1AU$ from the sun to calculate the expected polarization of those lines. This scenario corresponds to those secondary fluorescence lines from a comet near the perihelion. The results show that $P_{max}>10\%$ can be expected in most environments.

\section{Discussion and Summary}
We have studied the polarizations of primary and secondary fluorescence lines under the influence from GSA. These lines in optical and near infrared bands are observable with ground based telescopes. The effect of collisions are accounted for in our calculations.
The calculation for collision indicates that the collision  depolarize the ground and metastable states. When collision process is dominant over the pumping, the primary fluorescence from the excited states are synchronized to the perpendicular scattering, i.e., the polarization directions of the different scattered lines are the same orientation; the polarization from fine-structure levels  be zero. When the optical pumping is dominant, the polarization directions for different primary fluorescence lines are different from one-another. This difference is because the magnetic alignment on the ground state is transferred to the excited state. In this regime, the angle between different primary fluorescence polarization could provide constraint on the magnetic polar angles from the LOS ($\theta_B$).
The synergetic implementation of the primary fluorescence lines and meta-stable fine structure lines could reveal the 3D magnetic field locally. Earlier GSA studies have shown that the 3D magnetic field could be measured with fine structure lines by observing both the polarization angle and the degree of polarization (see, e.g., \citealt{YLhanle,YL12,ZY18}). Our new results have simplified the observational strategy for 3D magnetic field measurement because our work only requires the measurement of the polarization angles and could avoid the great observational challenge of precise measurement of polarization amplitude.

We have calculated the polarization produced from different environments, including interplanetary medium, ISM, and the QSO host galaxies. For general diffuse medium, measuring the polarization of atomic/ionic lines  provide us with the magnetic information because the magnetic alignment on the ground state  influence different atomic levels through pumping and fluorescence mechanisms. The calculations in our work show that the GSA is a measurable effect already for the optical and near-Infrared spectropolarimeter on the telescopes today, e.g., ESPaDOnS for 3.6-m Canada-France-Hawaii Telescope \citep{ESPaDOnSob}, HARPS for 3.6-m ESO telescope \citep{HARPS03}, PEPSI for Large Binocular Telescope \citep{PEPSI15}. When designing the observation with GSA, we could take into account the atomic lines in different wavebands. GSA observation with multi-wavelength atomic lines are advantageous because of the unique velocity information readily obtained with the spectroscopy, making 3D magnetic tomography achievable.

Our conclusions are:
\begin{itemize}
\item The polarization of the primary fluorescence lines could measure the magnetic polar angle compared to the LOS direction. Clear different behaviour can be observed between collision dominant regime and GSA regime.
\item The polarization of secondary fluorescence lines from metastable state are detectable in diffuse medium by UV/opitcal pumping with $P_{max}>10\%$ polarization for diffuse medium within and beyond our galaxy. The direction of the polarizations trace the 2D magnetic field in the plane of sky with a $90^{\circ}$-degeneracy.
\item 3D magnetic fields can be detected by the synergy of polarization directions of the primary and secondary fluorescence lines.
\item The polarization of fluorescence lines is a promising magnetic tracer with broad observational applications, for diffuse medium within and beyond our galaxy until early Universe.
\end{itemize}

\bibliography{yan}
\bibliographystyle{aasjournal}


\end{document}